\documentclass[11pt,preprint]{aastex}
\usepackage[]{natbib}
\usepackage[]{graphicx}
\usepackage[]{subfigure}

\newcommand\mdot{\rm \dot{M}}
\newcommand\msun{\rm M_{\odot}}
\newcommand\rsun{\rm R_{\odot}}

\newcommand\msunyr{\rm M_{\odot}\,yr^{-1}}
\newcommand\be{\begin{equation}}
\newcommand\en{\end{equation}}

\shorttitle{HBC722 NIR Variability}
\shortauthors{Green et al.}

\begin{document}

\title{Variability at the Edge: Optical Near/IR Rapid Cadence Monitoring of Newly Outbursting FU Orionis Object HBC 722}

\author{Joel D. Green\altaffilmark{1},
 	   Paul Robertson\altaffilmark{1},
            Giseon Baek\altaffilmark{2},
            David Pooley\altaffilmark{3,4},
            Soojong Pak\altaffilmark{1,2},
            Myungshin Im\altaffilmark{5},
            Jeong-Eun Lee\altaffilmark{2},
            Yiseul Jeon\altaffilmark{5},            
            Changsu Choi\altaffilmark{5},
            \& Stefano Meschiari\altaffilmark{1}
}

\affil{1. Department of Astronomy, 2515 Speedway, University of Texas at Austin, Austin, TX, USA \\
2. School of Space Research, Kyung Hee University, Gyeonggi-Do 446-741, Korea \\
3. Dept. of Physics, Sam Houston State University, Huntsville, TX, USA \\
4. Eureka Scientific, Austin, TX, USA \\
5. CEOU/Department of Physics \& Astronomy, Seoul National University, Seoul 151-742, Korea \\
}

\begin{abstract}

We present the detection of day-timescale periodic variability in the $r$-band lightcurve of newly outbursting FU Orionis-type object HBC 722, taken from $>$ 42 nights of observation with the CQUEAN instrument on the McDonald Observatory 2.1m telescope.  The optical/near-IR lightcurve of HBC 722 shows a complex array of periodic variability, clustering around 5.8 day (0.044 mag amplitude) and 1.28 day (0.016 mag amplitude) periods, after removal of overall baseline variation.  We attribute the unusual number of comparable strength signals to a phenomenon related to the temporary increase in accretion rate associated with FUors.  We consider semi-random ``flickering'', magnetic braking/field compression and rotational asymmetries in the disk instability region as potential sources of variability.  Assuming the 5.8 day period is due to stellar rotation and the 1.28 day period is indicative of Keplerian rotation at the inner radius of the accretion disk (at 2 $R_{\star}$), we derive a B-field strength of 2.2-2.7 kG, slightly larger than typical T Tauri stars. If instead the 5.8 day signal is from a disk asymmetry, the instability region has an outer radius of 5.4 $R_{\star}$, consistent with models of FUor disks.  Further exploration of the time domain in this complicated source and related objects will be key to understanding accretion processes.

\end{abstract}
\keywords{stars: pre-main sequence --- stars: variables: T Tauri,  stars: individual (HBC 722)}

\section{Introduction}

FU Orionis-type objects (hereafter, FUors) are a group of outbursting low-mass pre-main 
sequence objects.  The archetypal source, FU Ori, brightened by 6 magnitudes in B during
 a few month period in 1936.  About eight more flares are linked to sources observed prior to 2010.
 A second group of objects also generally considered FUors are characterized by similar spectral characteristics to the classical sources (broad blueshifted emission lines, IR excess, near-IR CO
overtone absorption; \citealt{hartmann96a, reipurth10}).  \citet{hartmann96a} theorized that FUors are the result of a sudden cataclysmic accretion of material from a gas reservoir accumulated in the circumstellar disk surrounding a young stellar object similar in evolutionary stage to a T Tauri star.  Models attributing the rise in system luminosity to the accretion process indicate 
that the accretion rate in an affected system rises over a few months from the typical rate for a T Tauri star ($\dot{M}$ $\lesssim$
10$^{-7}$ M$_{\odot}$ yr$^{-1}$) up to 
10$^{-5}\,$M$_{\odot}$ yr$^{-1}$, and then decays over $\sim$ 10-100 yr \citep{bell94}, although individual lightcurves show different decay profiles \citep{hartmann96a}.  
Over the entire duration of such an outburst the star might accrete 
$\sim$ 0.01 M$_{\odot}$ of material, roughly the mass of a typical T
Tauri disk \citep{andrews05}.  
FUors have some qualities that link them more closely to embedded sources than classical T Tauri stars, and some of them exhibit flat or rising mid-IR continuum indicative of a residual cold envelope.  FUors are thought to occur in repeating outburst cycles with replenished material from an infalling envelope \citep{weintraub91,sandell01,vorobyov05,green06, quanz07a,zhu10}, perhaps eroding their envelopes in stages.  Optical/near-IR SEDs of FUors are typically dominated by accretion luminosity rather than starlight; the optical peak of the SED is much broader than that of a single temperature blackbody and is dominated by the inner annuli of the disk, heated up to temperatures comparable to or exceeding the stellar surface during outburst (up to 9000 K).

There have been $\sim$ 10 FUors observed since 1936, a rate of one eruption per decade, but no outbursting object had followed the classical FUor profile in several decades.  Then in 2010, HBC 722 (also known as LkH$\alpha$ 188-G4, PTF10qpf, and V2493 Cyg) became a newly erupted FUor \citep{semkov10a,miller11}, located in the ``Gulf of Mexico'' in the southern region of the North American/Pelican Nebula \citep[520 pc distance;][]{straizys89, straizys93, laugalys06}.  HBC 722 suffers relatively low extinction, like many of the classical FUors, and was relatively well-characterized prior to eruption, providing a unique opportunity to study the phenomenon's effect in real-time.  SMA observations \citep{dunham12} indicate a fairly low maximum disk and envelope mass, suggesting HBC 722 is one of the most evolved FUors, and least luminous.

It has long been thought that accretion from a circumstellar disk to a central young stellar object occurs along lines of the central star's magnetic field, threading the inner disk.  It is hypothesized that the inner edge of the disk, from which accretion columns arise, is set by a balance between the ram pressure of the infalling material and the magnetic pressure from the dipole field of the central star.  In this picture, material rises along the field lines and is either ejected by stellar winds, jets, and outflows, or lands on the central star in a small number of  ``hot spots'' \citep[][and references therein]{lovelace05}.   As the star rotates, these spots must also rotate, and appear as slight periodic fluctuations in the lightcurve.  Observations of T Tauri stars \citep{stassun99,rucinski08} and Herbig Ae/Be stars \citep{rucinski10} find indications of the stellar rotation and asymmetry due to a collection of hot or cool spots \citep{herbst07}, but only sparse evidence of direct accretion processes.  Additionally, pulsations and random reddening events in young stellar objects were considered in \citet{kenyon00} but found to be unlikely as the cause of observed $\sim$ 0.03 mag variability.  Finally some evidence for hour-timescale variability has been found in short-lived excited modes in shorter-duration outbursts, EXors \citep{bastien11}, likely more evolved sources than FUors.

HBC 722 is a uniquely high accretion-rate young low-mass star with very little surrounding material.  The lightcurve in the optical and near-infrared may be dominated by accretion and disk luminosity rather than starlight.  The implications of this are threefold: first, the brightness of HBC 722 is determined by accretion processes in addition to the intrinsic stellar properties; second, the hotspots may be far more intense than in a T Tauri star; third, the rotation rate of the accreting columns may depend on the material entering the magnetosphere from outside, at a distinct rotational velocity from the star's rotation rate.

We present the $r$-band lightcurve (in the photometric system of the Sloan Digital Sky Survey; SDSS) of HBC 722 during April 2011 - July 2012, taken with the 2.1m Otto Struve Telescope at McDonald Observatory.   We analyze the variations in the lightcurve of HBC 722, and discuss the physical implications of the detected periodic variability in comparison to studies of older FUors and T Tauri stars.  We compare two possible pictures for the system, and consider the potential for future observations.

\section{Observations and data reduction}

HBC 722 (20:58:17.0 +43:53:42.9) was observed with CQUEAN (Camera for QUasars in EArly uNiverse; \citealt{kime11,park12}) installed as a visitor instrument on the 2.1m Otto Struve Telescope at McDonald Observatory.  CQUEAN is an optical CCD camera system optimized at 0.7-1.1 $\mu$m using a 1024 pixel deep-depletion CCD chip.  The camera, with a custom-made focal reducer attached, has a 4.7$\arcmin\times4.7\arcmin$ field-of-view and 0.276$\arcsec$ per pixel resolution.  Most of the observed photometry was taken in $r$-band (SDSS filter system; \citealt{fukugita96}).  Observations were taken during 42 nights in April, July, August, October, November, and December 2011, May and June 2012.  During the nights of 2011 July 4, August 30, November 1, 6, and 8, 2012 May 26, 29, and June 30, chosen for their photometric nighttime conditions, we observed HBC 722 for half-nights, using continuous rapid-cadence (minute timescale) exposures during the night.  On August 23 we observed for the full night.  On the other nights we took a small number of images per night to characterize longer-term variability.   A summary of the observations can be found in Table \ref{obslog}.

\begin{figure}
\begin{center}
\includegraphics[width=14.5cm, angle=0]{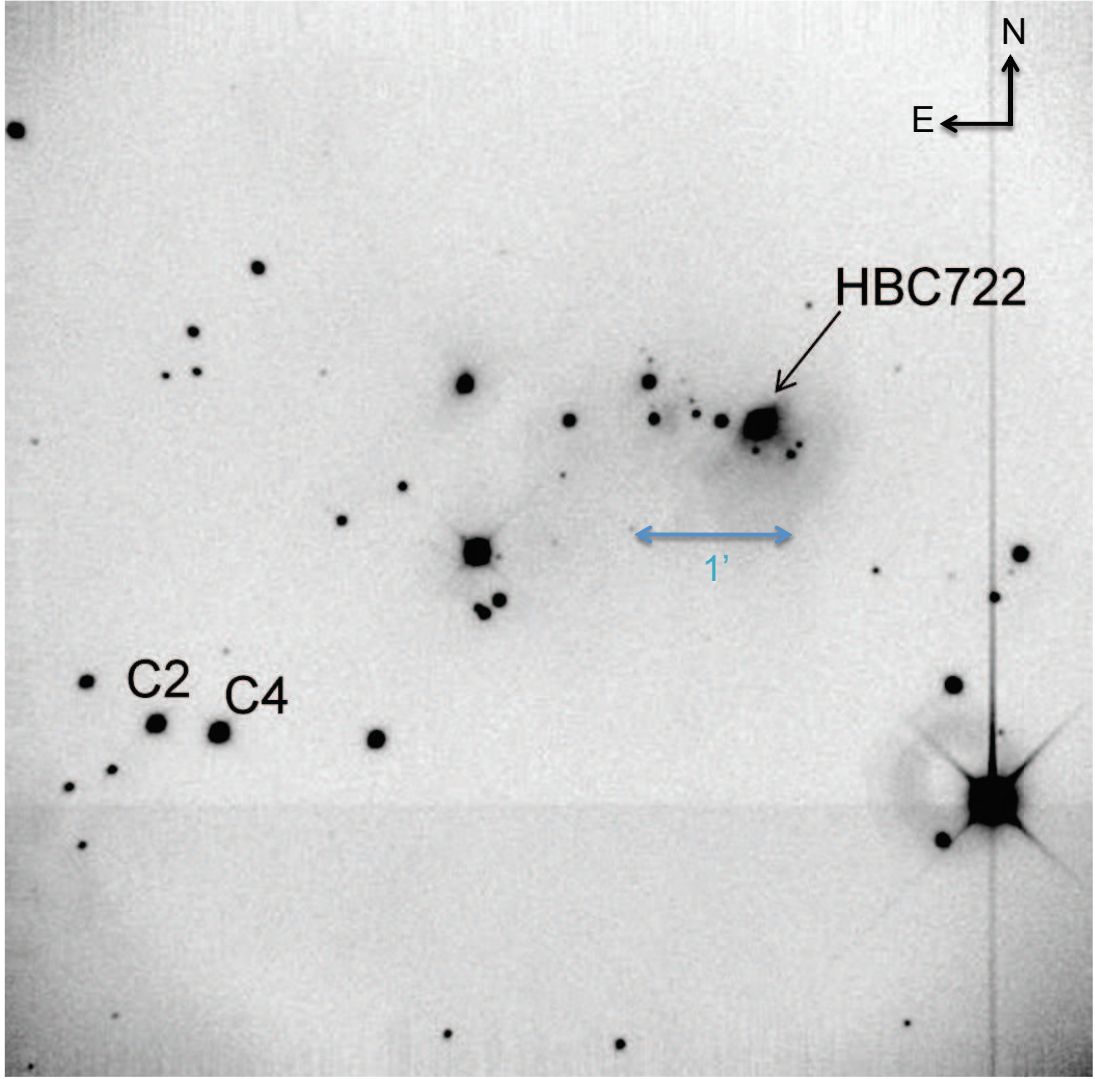}
\caption{$r$-band image of HBC 722 and surroundings (2011 Aug. 18).  The comparison stars used for differential photometry are labeled.}
\label{fov}
\end{center}
\end{figure}

All images were pre-processed with the IRAF\footnote{IRAF is distributed by the National Optical Astronomy Observatories, which is operated by the Association of Universities for Research in Astronomy, Inc., under cooperative agreement with the National Science Foundation.} packages.  We first subtracted CCD bias values from images taken before, during, and after observing the target, on each night.  Flatfielding utilized sky flat images obtained at both dawn and dusk when available.  Dark subtraction was used only for the images taken in December, because of short exposure time.  To perform aperture photometry, we used Source Extractor \citep{bertin96}; the size of the aperture was set by the FWHM observed seeing for that night.  We found that some of the background stars --themselves young stellar objects -- showed photometric variability, and carefully selected two background comparison stars (labeled as C2 and C4 in Figure \ref{fov}) not exhibiting significant variability in order to perform differential photometry.  C4 (1338-0391536; 20:58:30.04 +43:52:23.76 J2000; $r$-mag 14.41 $\pm$ 0.02) was used as a comparison star for the entire dataset; C2 (1338-0391541; 20:58:31.54 +43:52:28.56 J2000; $r$-mag 15.04 $\pm$ 0.02) was used as a check star to confirm the lack of variability in C4.  The RMS magnitude uncertainties are estimated from the comparison stars.

We utilize both high-cadence and isolated photometry from our observing runs.  The night of October 29 was not photometric, and has been removed from our data set.  Additionally, we removed 4 points (3 from July 10 and 1 from August 29) which were heavily offset from other data taken the same night.  
In addition to the full-night observations we present CQUEAN data for HBC 722 on 12 nights between 2011 August 18-31, 7 nights between 2011 Oct. 29-Nov. 8, 3 nights between 2012 May 26-30, and 7 nights between 2012 June 25-July 1, which we consider along with the rapid-cadence observations.

\section{Results}

\begin{figure}
\begin{center}
\includegraphics[width=14.5cm, angle=0]{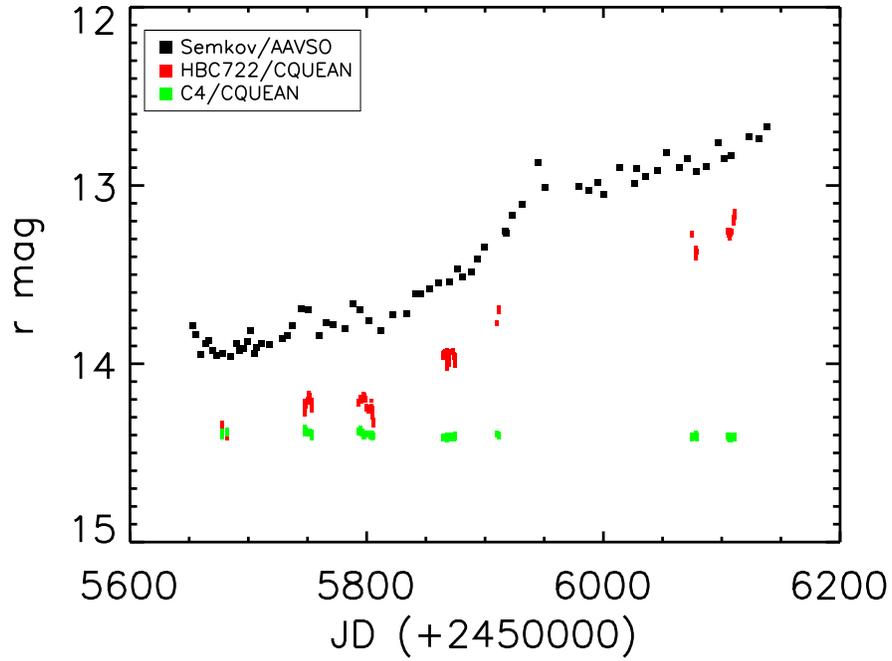}
\caption{$r$-band lightcurve for HBC 722 during our observation period (2011-12); the luminosity 
and time evolution observed by CQUEAN (red; AB-magnitude system) are consistent with catalog values (black; AAVSO/Johnson-Cousins system) \citep{semkov12a, semkov12b}.  We also present simultaneous observations of the comparison star C4 (green), which remains stable over time. 
}
\label{lightcurve}
\end{center}
\end{figure}

After initially decaying from peak outburst between September 2010 and April 2011, HBC 722 remained relatively constant in luminosity until September 2011.  After, HBC 722 displayed a gradual brightening; this is 
consistent with the AAVSO observations from the same time period (Figure \ref{lightcurve}).  Our observations cluster in the latter two epochs.  To fit short-period variation, we accounted for this slow brightness increase by treating our five observing runs as separate datasets and including constant magnitude offsets for each dataset as free parameters in our lightcurve fit.  In addition to the intrinsic behavior of the star, the offsets prevent any long-period photometric baseline drifts caused by the instrument from influencing our results.  Because we did not take sufficient data in April or December to properly establish this constant offset, we have excluded those data from our model.  As a check on the accuracy of our differential photometry, we also present results for comparison star C4, which remains steady in brightness of this same time period (Figure \ref{lightcurve}).

\begin{figure}
\begin{center}
\includegraphics[width=14.5cm, angle=0]{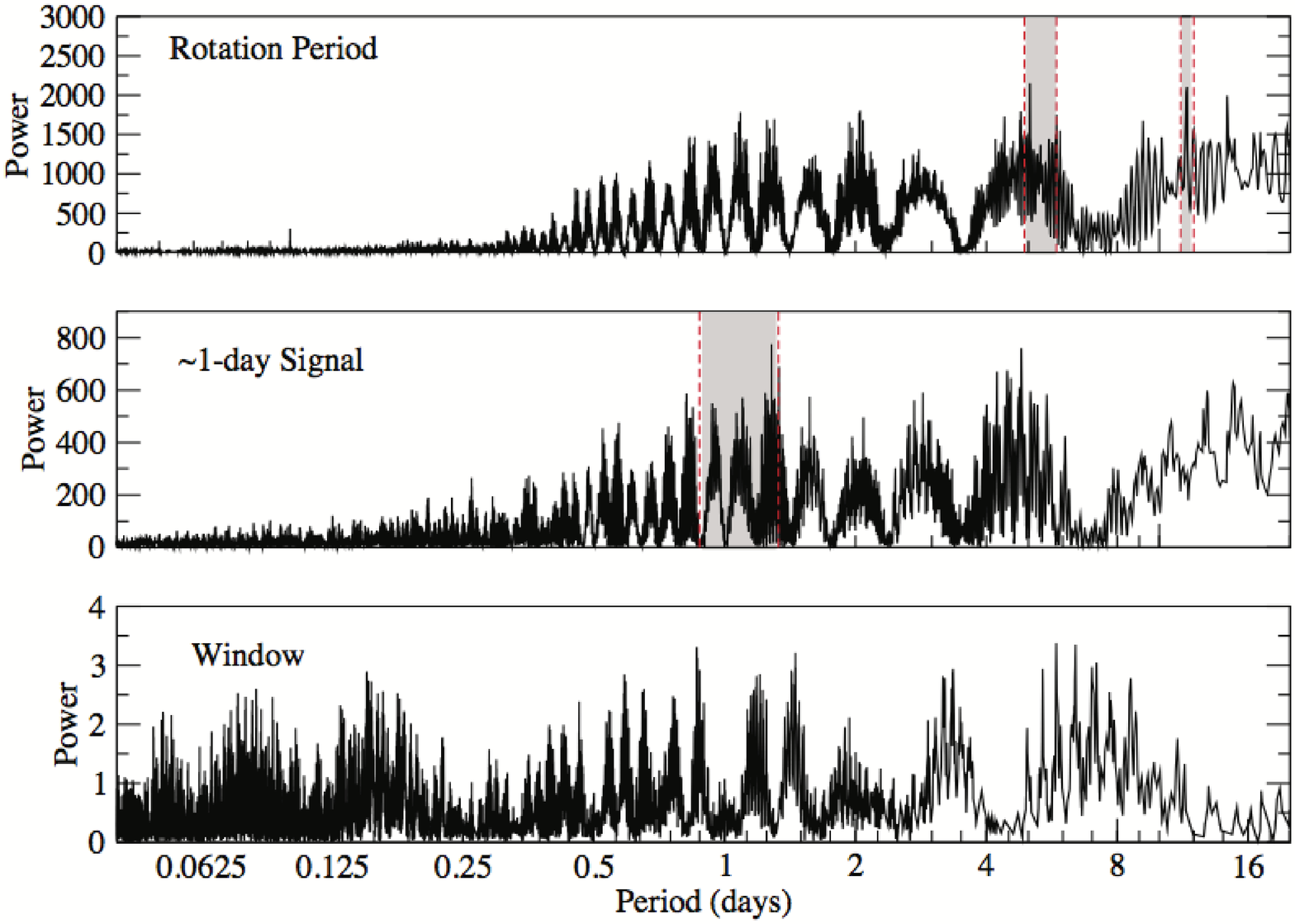}
\includegraphics[width=14.5cm, angle=0]{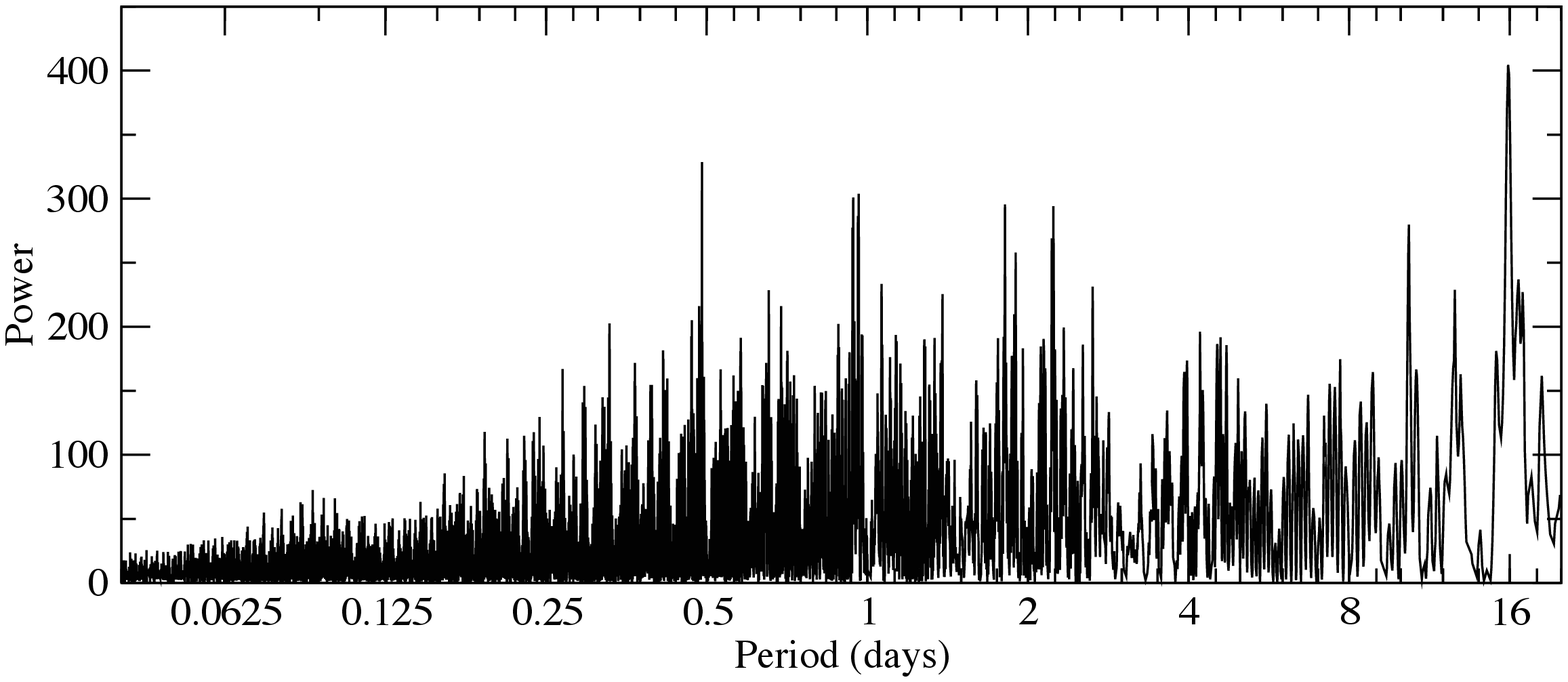}
\caption{Generalized Lomb-Scargle periodogram of HBC 722 (top) and comparison star C4 (bottom).  From the top down, we show the periodograms of the raw lightcurve, the residuals after subtracting the rotation period, and the window function, respectively, for HBC 722.  The bottommost periodogram is the raw lightcurve for C4.  The shaded region in the top plot represents the range of acceptable fits to the data; the shaded region in the middle plot represents the resulting range in the shorter period from the choice of fits to the longer period.}
\label{periodogram}
\end{center}
\end{figure}

We then computed a generalized Lomb-Scargle periodogram \citep{zechmeister09} from the  photometry, which we present in Figure \ref{periodogram}.  We chose this periodogram algorithm  because of its versatility in identifying periodic signals of both stellar and planetary nature, for unevenly sampled data \citep[e.g.][]{kurster03,robertson12a,robertson12b}.  We also applied this technique to the comparison star, C4.  We do not consider periods greater than 16 days; our data lack sufficient time domain coverage and offset mismatches between observations may masquerade as longer periods.  The resulting periodogram shows significant power most strongly in bands near $\sim$ 5 and 12 days.  The strongest peak is found at 5.8 days.

\begin{figure}
\begin{center}
\subfigure[\label{rotation}]{\includegraphics[scale=0.45]{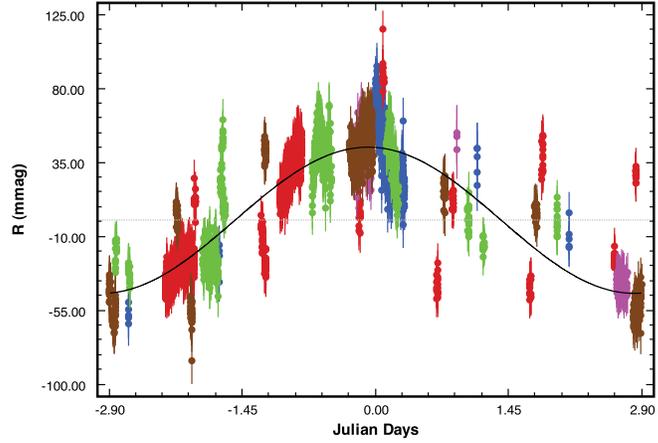}}
\subfigure[\label{stream}]{\includegraphics[scale=0.45]{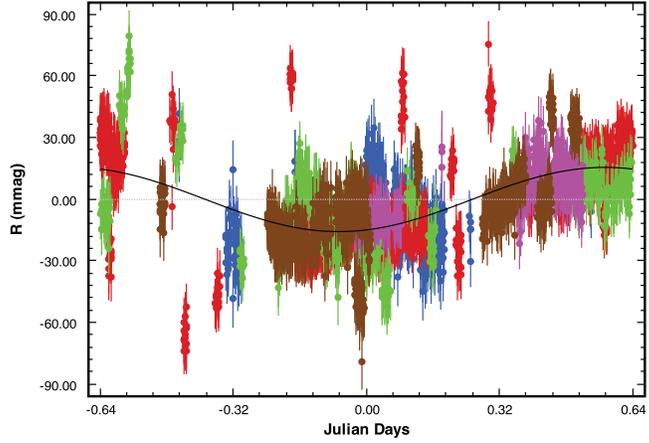}}
\subfigure[\label{alias}]{\includegraphics[scale=0.45]{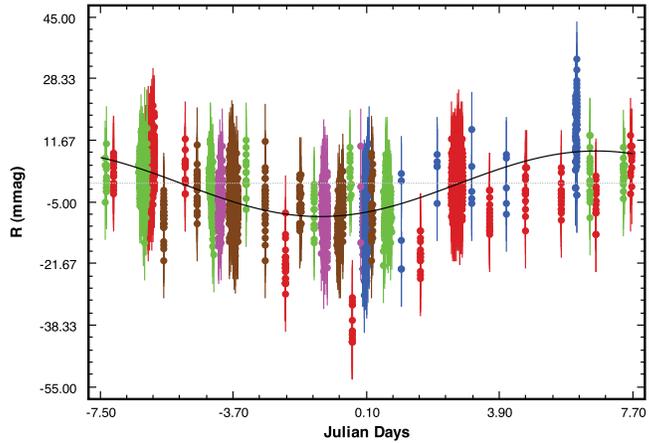}}
\caption{{\bf Top and Middle:} Phase plots of the 2 signals in our final lightcurve model for HBC 722.  The blue, red, green, pink, and brown points represent the photometry from our 2011 July, 2011 August, 2011 October/November, 2012 May, and 2012 June observing runs, respectively.  We identify the signals as \emph{a}. the 5.8 day stellar rotation period and \emph{b}. an inner disk/accretion-related phenomenon with a 1.28 day orbital period. \emph{c}. presents the strongest signal observed in comparison C4, at 15.8 days.}
\label{model}
\end{center}
\end{figure}

We fit a sinusoidal model to the lightcurve using the SYSTEMIC console \citep{meschiari09}, which uses multidimensional Levenberg-Marquardt minimization to optimize $\chi^2$ over all fitted parameters at once.  There are several solutions of similar statistical significance, but they converge into two families of periods.  We found convergence to a sinusoidal fit at $P = 5.8$ days.  We show the lightcurve folded to the 5.8 day period in Figure \ref{rotation}.  As seen in the second  panel of Figure \ref{periodogram}, the power spectrum of the residuals to the 5.8 day period shows a group of signals near 0.9 and 1.3 days, with a distinct peak at $\sim$ 1.28 days (Figure \ref{stream}).  As with the 5.8 day peak, a ``bootstrap'' false alarm probabiliity (FAP) estimate gives FAP $< 10^{-3}$.  Using SYSTEMIC we fit a sinusoid to the data, finding convergence at $P = 1.28$ days.  Our final 2-signal fit gives periods of 5.8 and 1.28 days with amplitudes of 44.4 $\pm$ 0.5 mmag and $15.6 \pm 0.4$ mmag, respectively, in the $r$ band. 

\subsection{Statistical significance and comparison to C4}

Although our formal period errors 
to these specific fits are minuscule, a realistic uncertainty in the fitted period is marked in the bands in Figure \ref{periodogram}.  We considered two different benchmarks to verify the significance of the spectrum peak.  First, we have run a Monte Carlo resampling of our data, performing a bootstrap FAP estimate equivalent to that of \citet{kurster97}.   The bootstrap approach retains the original timestamps of the dataset, and assigns a value from the photometry at random to each timestamp.  We created 1000 such noise data sets, and the periodogram power of those sets never exceeded the power in the 5.8-day peak, placing an upper limit of $10^{-3}$ on the FAP.  Second, we perform the same analysis on C4.

The phase-folded signals for HBC 722 and C4 are shown in Figure \ref{model}.  The model for HBC 722 gives a reduced $\chi^2$ of 1.64 and an rms scatter of $0.015$ $r$ mag.  Because HBC 722 is an active star, we anticipate a certain amount of high-frequency noise in our data, which is likely causing our fitting statistics to be somewhat higher than ideal.  Furthermore, we do in fact see additional power in the periodogram of the residuals to a 2-signal fit, but since the signals are at very short periods, and have amplitudes smaller than the average internal error of our individual data, we have not considered these signals in detail.

The phase-folded signal for C4 indicates a weak periodicity, with an amplitude of 8.86 mmag at $r$ band, and  a period of 15.77 days.  Although this periodicity may be real and attributable to rotation or accretion processes in C4, we do not consider signals at this significance level, and any variability in C4 should have little effect on our analysis of HBC 722.

\section{Discussion}

\subsection{Comparison to other studies of young stellar objects}

From the previous analysis, we uncover two robust period families, around 5.8 and 1.28 days, with $r$ band amplitudes of 44 and 16 mmag.   FU Ori, the prototype of the class, outburst in 1936 and has remained close to peak brightness since.  \citet{kenyon00} modeled the lightcurve as fading at a rate of only 14 mmag yr$^{-1}$ at $B$-band.  FU Ori is at a comparable distance and has a similar estimated central mass to HBC 722 and is thus the closest comparison, although the HBC 722 flare is less in both luminosity increase and implied accretion rate.

For FU Ori, \citet{kenyon00} find a non-periodic variability of 33 mmag at $V$ band, after removal of a 17-day period with an amplitude of 9 mmag.  They use a lower sampling rate ($\sim$ few per day) but consider a multiband (UBVR) dataset, and conclude that 35 mmag variations (correlated over multiple color bands) must occur on timescales $<$ 1 day, comparable to the dynamic timescale of the inner disk.  They point out that if associated with rotation, 1-2 day periods for FU Ori put the star at close to breakup, as would the 1.28 day period we detect in HBC 722.  They conclude that ``flickering'' (due to dynamics of the inner disk) is the most likely cause of short-term variability, rather than stellar rotation, magnetospheric accretion, or other causes.  Unlike FU Ori, HBC 722 exhibits a much stronger (44 mmag) period at 5.8 days.

The results of \citet{clarke05} provide a higher sampling rate comparison to HBC 722.  Using UBVR photometry, they find periods of $\sim$ 14 days for both V1057 Cyg and V1515 Cyg, comparable to the 17 day period found in FU Ori.  In addition, non-periodic oscillations are noted at shorter timescales.  Interestingly, V1057 Cyg and V1515 Cyg are fading much more quickly than FU Ori.  In HBC 722, the $r$ band data hints at some periodicity on $\sim$ 12 day timescales but the power spectrum peaks at shorter periods.  They note that any day-timescale periodic variability ($\sim$ 0.1 mag) in the older FUors is not persistent over $>$ 1 yr timescales, and attribute the non-periodic variations to flickering.

EXors are often thought to be more evolved versions of FUors, exhibiting smaller but more rapid variability, attributed to disk instabilities in otherwise classical T Tauri-like systems.  Two well-studied EXors include the prototype EX Lupi, and V1647 Ori.  Both have both undergone multiple flares in the past few decades, of shorter duration and relaxation timescale than FUor flares.  \citet{bastien11} consider a lightcurve at a similar sampling rate to ours, for the EXor V1647 Ori.  They find very rapid timescale variation (0.13 days, at 51 mmag amplitude in $z$ band) during the  brightening event in 2003.  They also conclude that this rapid timescale variability cannot be attributed to stellar rotation, and associate variability with ``flicker noise''.  This is comparable in some ways to HBC 722, which rebounded from initial post-outburst decay to brighten during a second epoch throughout our observations.  \citet{bastien11} do not observe the 0.13 day period during the 2009 brightening phase of V1647 Ori, suggesting that the effect was transitory.  We cannot eliminate the possibility of $\sim$ 0.1 day periods in HBC 722; although we detect periods of lower amplitude on comparable timescales to 0.1 day, the statistical significance of our data is not yet sufficient to claim detection.  

Finally, \citet{stassun99} detect periodicities in T Tauri stars attributed to stellar rotation, typically between 1 and 10 days, although they range down to 0.1 days in some cases, using sparser time sampling.

\subsection{Two Interpretations of Periodicity}

We can interpret the two signal families in our lightcurve in a few different ways, presuming they are stable on $>$ 1 yr timescales.  In one picture, the 5.8 day signal is the rotation period of HBC 722, consistent with typical T Tauri stellar rotation periods \citep{stassun99}, and FUors \citep{kenyon00,clarke05}.  The 1.28 day period is then attributable to the infalling material.  This case requires considerable braking of material to allow accretion: the infalling material would be rotating at a rate 4.5 times faster than the stellar magnetic field at the inferred radius (see below).

In a second picture, we consider the reverse interpretation, with the 5.8 day period attributed to the infalling material, and the star rotating at 1.28 days.  The disk instability/infall region in 2D models of the disk of FU Ori extends out to 0.5 AU \citep{zhu07}.  FU Ori has an accretion rate of 2 $\times$ 10$^{-4}$ $\msun$ in these models, about 100 times greater than the inferred rate for HBC 722 (see below).  The instability region in HBC 722 could be 10 times smaller in radius and maintain this lower accretion rate, at the same densities.

In this picture, the 5.8 day period could be attributed to a rotating asymmetry in the instability region of the disk.  The radius of the asymmetry, and thus the infalling material, would be at 5.4 $R_{\star}$, for Keplerian rotation, broadly comparable to the size of the instability region scaled down from the model for FU Ori.  This radius may move inward, or the asymmetry could dissipate, as material drains from this region.  If the 1.28 day period is attributable to stellar rotation -- faster than typical for a T Tauri star but not outside of the normal distribution \citep{stassun99} -- then the central star must be close to or at breakup velocity.  The magnetospheric radius would still be at 2.0 $R_{\star}$; thus the area of the disk instability would range from 2.0 to 5.4 $R_{\star}$. 

Both physical models have difficulties.  The first case requires strong braking at the magnetospheric radius.  In this picture, the slowly-rotating magnetic field must do a considerable amount of braking.  The angular velocity of material with a period of 1.28 days would be much greater than the stellar magnetospheric rotation period of 5.8 days.  Material that is not sufficiently slowed by the infall process would not land on the central star.  Models of FUor disk instabilities or T Tauri gas disks generally assume that the inner edge of the gas disk determined by the dust destruction radius \citep[e.g.]{kamp04,zhu07}.  Once the dust opacity is removed, the gas rapidly heats and evaporates, truncating the disk.  Thus it seems implausible that the disk could extend all the way to the central star, and material rotating at Keplerian velocities at the radius indicated by a 1.28 day period would not accrete onto HBC 722.  The second case requires a very rapidly rotating central star and a higher amplitude asymmetry maintained over a year timescale at the {\it outer} radius of the instability, rather than the inner edge.

\subsection{Accretion-related variability}

The inner edge of the disk in a T Tauri star is expected to be magnetically locked to the stellar rotation by magnetic fields passing through the magnetospheric radius (hereafter, $R_{\rm mag}$) in the disk.  The bulk of the disk should be rotating at Keplerian velocity; thus the disk inner edge is the point at which the magnetic field begins to ``brake'' the material.  However, in the case of a burst accretion event, the accretion rate rises and the inner disk edge may compress the magnetic field, or interact with the field in an interference pattern.  Over time the inner disk would either accrete onto the central star, become ejected, or equilibrate with the stellar rotation.  

If we hypothesize that the 1.28 day (2.0 $R_{\star}$ for Keplerian rotation) period is the most reliable indicator of the accreting inner disk edge (in both scenarios presented above), then by making some simple assumptions we can calculate a number of physical properties of the star and disk, and determine if they deviate significantly from those of a typical T Tauri star.  We follow a simplified procedure comparable to that used in models of FU Ori \citep{shu07,konigl11}.

The luminosity had decreased to 5.4 L$_{\odot}$ \citep{kospal11,green11b} by March 2011, down from the peak luminosity of 12 L$_{\odot}$ in September 2010.  As of mid-2012 it has since returned to peak luminosity.  Pre-outburst HBC 722 was a K7-M0 emission line object \citep{cohen79}, a T Tauri star.  Assuming that the current luminosity is entirely attributed to accretion, we can determine the accretion rate \citep[e.g.][]{hartmann98}:

\begin{equation}
L_{\rm acc} =  \frac{GM\mdot}{2R_{\star}}
\end{equation}

Assuming $R_{\star}$ $=$ 2.0 R$_{\odot}$, and a central mass of 0.5 $\msun$ from pre-outburst spectral type measurements, we calculate the accretion rate $\mdot$ $=$ 1.31 $\times$ 10$^{-6}$ $\msunyr$ \citep[consistent with previous estimates;][]{kospal11} and Keplerian orbital radius r$_{kep}$ $=$ 2.0 $R_{\star}$ for the detected period of 1.28 days.

An upper limit to the rate of angular momentum injected by the infalling stream onto the star would be

\begin{equation}
\frac{dl_{\rm acc} }{dt} = M_\star  v_{kep} r_{kep}
\end{equation}

where v$_{kep}$ is the Keplerian rotation rate at r$_{kep}$, assuming all of the material lands on the central star and is not ejected.  The angular momentum transferred by material landing on the star over 1 yr would increase the star's angular momentum by only one part in $\sim 0.5 \times 10^{5}$ (a decrease in the rotation rate of $\sim$ 1 sec/yr for a 5.8 day period).  Thus over an entire outburst ($\sim$ 100 yr) we might expect the star to spin up by one part in 10$^4$, not easily detectable using this technique, even without accounting for magnetic braking which would counteract this effect.  

\subsubsection{Dipole Magnetic Field Strength}

\citet{krull07} measured the magnetic field strength in T Tauri stars from Zeeman broadening of spectral lines.  Zeeman measurements for HBC 722 would help discriminate whether magnetosheric accretion is a plausible contributor to variability.  We have proposed for, but not yet acquired, spectroscopic data for Zeeman measurements in HBC 722.  In our current picture, we can predict the magnetic field strength, tied to the inner disk edge.  The magnetospheric radius is the location in which the stellar dipole field intersects the disk, and we can calculate the magnetic field strength independently.

\begin{equation}
B = \frac{\mu}{R^3}
\end{equation}

where $B$ is the dipole magnetic field of the star, $\mu$ is the dipole moment, and $R$ is the stellar radius, in CGS units.

The disk truncation should occur where the ram pressure of the infalling material and the magnetic energy balance, at the magnetospheric radius $R_{\rm mag}$.  We can relate $R_{\rm mag}$ to the magnetic moment using \citet[][Eq. 1]{patterson94}.  As noted therein, for a disk the magnetospheric radius is $\sim$ 0.5 $R_A$, the Alfven radius \citep{lamb88,konigl11}.

We then derive magnetic field strengths at the magnetospheric radius of 2.2-2.7 kG for central stellar masses of 0.45 - 0.6 $\msun$.  This is consistent with results from \citet{krull07}, if somewhat larger than typical.  They derive magnetic field strengths of $\sim$ 1 kG around classical T Tauri stars, and around FU Ori \citep{donati05}.  Although the periodicity is stable in some form over a full year timescale, the presence of a ``family'' of periods may be suggestive of ``flickering'' rather than a fixed asymmetry at the inner disk edge.

\subsection{Extent of the Inner Disk}

It has been noted that the disappearance of lines such as Br$\gamma$ correlates with the ignition of outbursts and may indicate a compression of the magnetic field toward the surface of the star \citep{benisty10}.  If the magnetospheric radius was compressed from, for example, 4.7 to 2.9 $\rsun$, or 8.6 to 7.4 $\rsun$, the difference in period would be of the order of 1.28 days, resulting in an ``interference pattern''.  The system luminosity began to increase during the course of our  observations; thus we may be probing a variation in the magnetospheric radius, an interaction with the disk edge.  If the magnetosphere is compressed, we may be seeing the interaction between the new fast-rotating disk edge and the slower rotating stellar surface.  Alternately an asymmetry at the very edge of the disk could cause this effect.

\section{Conclusions}

We find families of periodicity at 5.8 and 1.28 days, with amplitudes of 44 and 15.6 mmag in $r$, respectively.  In one picture, we attribute the 5.8 day period to stellar rotation, and the 1.28 day period to Keplerian rotation at the magnetospheric radius, the inner disk edge.  In a second picture, we reverse this interpretation HBC 722 appears to be a typical T Tauri star ringing from a dramatic accretion event.  Under these assumptions, and presuming the central stellar mass to be 0.5 $\msun$, we derive an inner disk radius of 2.0 $R_{\star}$, and a magnetic field strength of 2.2 - 2.7 kG, consistent with the characteristics of HBC 722 from pre-outburst observations, but with a greatly enhanced accretion rate.  In the latter case, we derive a disk instability region ranging from 2.0 to 5.4 $R_{\star}$. If correct, it appears that the magnetic field in a recent FUor outburst does not differ significantly from quiescent T Tauri stars, perhaps slightly enhanced or compressed.  Large scale surveys of this nature may unlock accretion properties, through rapid-cadence photometric monitoring campaigns of quiescent and  variable stars.  Multiband color observations will provide additional constraints through the temperature and line profiles will track the variability of different velocity components.  If the variability in HBC 722 is attributable to the accretion process, greatly enhanced by the event that triggered the outburst, then recently outbursting FUors are ideal laboratories to study accretion theory using lightcurve variability.

\acknowledgements

The authors would like to thank Sally Dodson-Robinson, Isa Oliveira, Augusto Carballido, Dan Watson, Dan Jaffe, Neal Evans, John Lacy, John Barentine, Bill Cochran, Mike Pavel, and Shakthi Shrima for helpful discussions.  PR is supported by a University Continuing Fellowship.  GB, SP, MI, YJ, and CC acknowledge the support from the Creative
Research Initiative program, No. 2010-0000712, of the National Research
Foundation of Korea (NRFK) funded by the Korea government (MEST).  The research of J.-E. L. is supported by Basic Science Research Program through the National Research
Foundation of Korea (NRF) funded by the Ministry of Education, Science and Technology (No.
2012-0002330).  S.M. acknowledges support from the W. J. McDonald Postdoctoral Fellowship.  This paper includes data taken at The McDonald Observatory of The University of Texas at Austin.


\begin{center}
\begin{deluxetable}{l r r r r}
\tabletypesize{\scriptsize}
\tablecaption{CQUEAN Monitoring of HBC 722 Observing Log \label{obslog}}
\tablewidth{0pt}
\tablehead{
\colhead{Date(UT)} &  \colhead{Time(UT)} & \colhead{Exposure Time(sec)} 
& \colhead{Number of frames}}
\startdata
2011-04-26 &	10:51:58	 &		20 &	25	\\
2011-04-30 &	10:59:49	 &		20 &	30	\\
2011-07-05 &	05:44:11	 &		10 &	312 \\
2011-07-06 &	08:23:28	 &		10 &	5	\\
2011-07-07 &	08:28:23	 &		10 &	5	\\
2011-07-08 &	08:06:10	 &		10 &	5	\\
2011-07-09 &	08:49:03 &		10 &	5	\\
2011-07-11 &	07:32:59 &		10 &	59 \\
2011-08-19 &	08:25:06 &		20 &	25 \\
2011-08-21 &	06:23:16 &		20 &	15 \\	
2011-08-22 &	06:37:40 &		20 &	15 \\	
2011-08-23 &	04:37:15 &		20 &	15 \\	
2011-08-24 &	02:35:07 &		20 &	112 \\
2011-08-25 &	04:36:04 &		20 &	15 \\	
2011-08-26 &	05:11:27 &		20 &	15 \\	
2011-08-27 &	05:29:06 &		20 &	15 \\	
2011-08-28 &	04:54:20 &		20 &	15 \\	
2011-08-29 &	05:16:30 &		20 &	15 \\	
2011-08-30 &	05:02:21 &		20 &	18 \\	
2011-08-31 &	03:34:55 &		20 &	80 \\
2011-09-01 &	06:18:15 &		30 &	15 \\	
2011-10-30 &	03:12:20 &		20 &	15 \\	
2011-10-31 &	02:28:49 &		20 &	15 \\	
2011-11-01 &	02:12:01 &		20 &	15 \\	
2011-11-02 &	01:05:49 &		20, 30 &	12, 204 \\
2011-11-04 &	01:07:04 &		30 &	120 \\
2011-11-05 &	02:14:58 &		30 &	15 \\	
2011-11-07 &	00:47:35 &	         30 &	15 \\	
2011-11-08 &	01:31:37 &		30, 60 &	3,12 \\	
2011-11-09 &	00:48:21 &		30 &	165 \\
2011-12-14 &	02:09:20 &		90 &	2 \\	
2011-12-16 &	01:16:41 &		90 &	11 \\
2012-05-27 &   08:28:31 &                  20, 30 & 45, 85 \\
2012-05-30 &   07:24:56 &                  15 & 299 \\
2012-05-31 &   09:51:59 &                  10 & 3 \\
2012-06-26 &   07:54:30 &                  15 & 30 \\
2012-06-27 &   06:58:05 &                  15 & 29 \\
2012-06-28 &   05:09:22 &                  15, 20 & 221, 177 \\
2012-06-29 &   05:43:15 &                  15 & 30 \\
2012-06-30 &   05:26:16 &                  30 & 30 \\
2012-07-01 &   07:10:07 &                  30 & 183 \\
2012-07-02 &   06:33:15 &                  15, 30 & 3, 30 \\
\enddata
\tablecomments{Observing Log of CQUEAN monitoring of HBC 722.  All observations listed are $r$-band and targeted at HBC 722.}
\end{deluxetable}
\end{center}

\end{document}